\documentclass[journal=jacsat,manuscript=article]{achemso}

\usepackage[version=3]{mhchem} 
\usepackage[dvipsnames]{xcolor}
\usepackage{hyperref}

\UseRawInputEncoding

\author{Chi Ian Jess Ip}
\affiliation{Pritzker School of Molecular Engineering, University of Chicago, Chicago, IL 60637, USA}
\alsoaffiliation{Current address: Department of Physics, Massachusetts Institute of Technology, Cambridge, MA 02139, USA}
\altaffiliation{These authors contributed equally to this work.}
\author{Qiang Gao}
\affiliation{Pritzker School of Molecular Engineering, University of Chicago, Chicago, IL 60637, USA}
\altaffiliation{These authors contributed equally to this work.}
\author{Khanh Duy Nguyen}
\author{Chenhui Yan}
\author{Gangbin Yan}
\author{Eli Hoenig}
\author{Thomas S. Marchese}
\affiliation{Pritzker School of Molecular Engineering, University of Chicago, Chicago, IL 60637, USA}
\author{Minghao Zhang}
\affiliation{Department of NanoEngineering, University of California San Diego, La Jolla, CA 92093, USA}
\author{Woojoo Lee}
\author{Hossein Rokni}
\affiliation{Pritzker School of Molecular Engineering, University of Chicago, Chicago, IL 60637, USA}
\author{Ying Shirley Meng}
\affiliation{Pritzker School of Molecular Engineering, University of Chicago, Chicago, IL 60637, USA}
\alsoaffiliation{Department of NanoEngineering, University of California San Diego, La Jolla, CA 92093, USA}
\author{Chong Liu}
\author{Shuolong Yang}
\affiliation{Pritzker School of Molecular Engineering, University of Chicago, Chicago, IL 60637, USA}
\email{yangsl@uchicago.edu}

\title[An \textsf{achemso} demo]
  {Preservation of Topological Surface States in Millimeter-Scale Transferred Membranes}

\abbreviations{TI,ARPES}
\keywords{Topological insulator, Bi_{2}Se_{3} Film, millimeter-scale transfer, ARPES}

\begin{document}






\begin{abstract}
 Ultrathin topological insulator membranes are building blocks of exotic quantum matter. However, traditional epitaxy of these materials does not facilitate stacking in arbitrary orders, while mechanical exfoliation from bulk crystals is also challenging due to the non-negligible interlayer coupling therein. Here we liberate millimeter-scale films of topological insulator Bi$_2$Se$_3$, grown by molecular beam epitaxy, down to 3 quintuple layers. We characterize the preservation of the topological surface states and quantum well states in transferred Bi$_{2}$Se$_{3}$ films using angle-resolved photoemission spectroscopy. Leveraging the photon-energy-dependent surface sensitivity, the photoemission spectra taken with $6$~eV and $21.2$~eV photons reveal a transfer-induced migration of the topological surface states from the top to the inner layers. By establishing clear electronic structures of the transferred films and unveiling the wavefunction relocation of the topological surface states, our work paves the physics foundation crucial for the future fabrication of artificially stacked topological materials with single-layer precision.
\end{abstract}

Keywords: Topological insulator, Bi$_{2}$Se$_{3}$ film, millimeter-scale transfer, ARPES

Van der Waals (vdW) materials, exemplified by graphene and transition metal dichalcogenides, provide versatile material platforms to realize fascinating topological \cite{FeiZaiyao2017NP, ESpanton2018Science, ChenGR2020, SerlinM2020Science, LiTingxin2021Nature} and strongly correlated phases of matter \cite{YeJT2012Science, Andras2013NM, JHPablo2018YCao1, JHPablo2018YCao2, Wanglei2020}. The vdW interactions between the 2D layers enable convenient mechanical exfoliation and re-assembly of new heterostructures. However, it is notably challenging to employ traditional exfoliation to produce ultrathin topological insulator (TI) films, such as Bi$_2$Se$_3$ \cite{HSteinberg2010, JGC2011, SMichael2011SCho} and Bi$_2$Te$_3$ \cite{KMFShahil2010, DTeweldebrhan2010}, with uniform thicknesses. This is due to the stronger orbital overlap between the outer chalcogen atoms in the TI films. In a recently developed approach termed ``remote epitaxy", thin films are fabricated on top of graphene-coated substrates followed by releasing of the epitaxial layers via mechanical exfoliation\cite{kim_remote_2017}. However, this method requires delicate control of the graphene buffer layer and the graphene-substrate interface, and has not been implemented on a wide variety of materials such as the TIs. Another exfoliation protocol employs gold coating on silicon to exfoliate and transfer Bi$_2$Se$_3$, but this approach does not enable accurate control of the film thickness over a large spatial scale \cite{TaoLin2021Nanoscale}. If one can transfer TI membranes with precisely controlled thicknesses at spatial scales larger than 1 mm, many revenue opportunities such as the construction of superlattices, devices, twistronics, and suspended films will be opened up, enabling new approaches to electronics, spintronics, and quantum informatics.

By etching the oxide layer at the oxide-chalcogenide interface, Bansal et al. separated molecular beam epitaxy (MBE)-grown Bi$_2$Se$_3$ thin films from oxide substrates \cite{NBansal2014}. This pioneering study highlighted the possibility of assembling new heterostructures using MBE-grown, wafer-scale films, yet it remains to be seen whether the electronic structure is preserved after the harsh chemical processing. In particular, it is unclear whether the key electronic state for TI films, the topological surface state (TSS), survives the chemical treatment. In this work, we employ angle-resolved photoemission spectroscopy (ARPES) to characterize a series of Bi$_2$Se$_3$ ultrathin films with thicknesses ranging from 3 to 10 quintuple layers (QLs), and reveal the electronic structure changes after a millimeter-scale liberation. Our ARPES data using $6$~eV photons suggest that despite contamination-induced spectral broadening, the key electronic structures including the TSS and the quantum well states (QWSs) are largely preserved after transferring the entire Bi$_2$Se$_3$ film from an SrTiO$_3$ substrate to a Si substrate. This result establishes the material physics foundation for further manipulation of such TI films for twistronics and topotronics. Interestingly, our ARPES spectra taken with $21.2$~eV photons on the same set of samples display no signatures of the TSS and diffusive features of other electronic bands. Considering the extreme surface sensitivity of $21.2$~eV photoemission~\cite{seah1979}, our work reveals the wavefunction relocation of the TSS from the top layer to the inner layers. This discovery has direct implications on the topology of the liberated TI films and the functionalities of future devices utilizing them.

\begin{figure}
\includegraphics[width=1\textwidth]{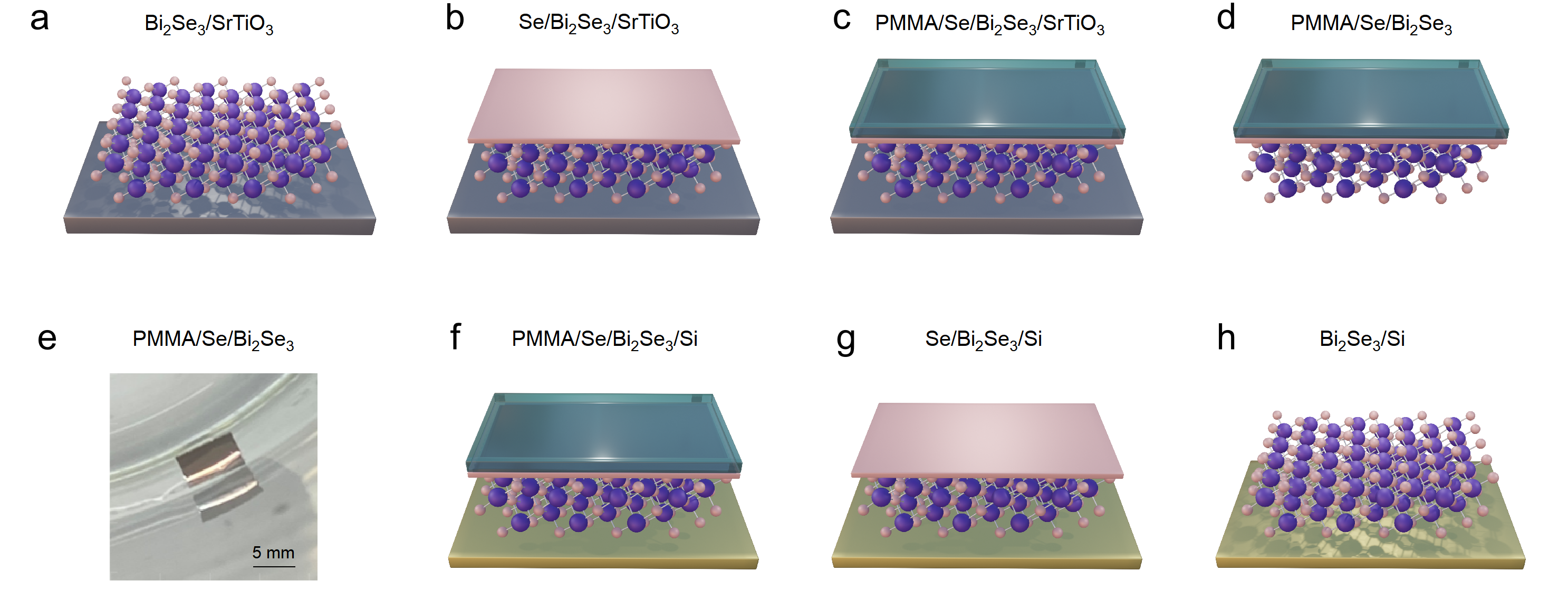}
  
  \caption{Schematics of the liberation process. (a) A Bi$_2$Se$_3$ thin film grown on an oxide substrate SrTiO$_3$ by MBE. (b) An amorphous selenium capping layer was deposited on the sample. (c) A layer of polymethyl methacrylate (PMMA) was spin-coated on top of the sample. (d) The Bi$_2$Se$_3$ thin film supported by PMMA was liberated from SrTiO$_3$ using buffered oxide etchant. (e) Photograph of a liberated $6$~QL Bi$_{2}$Se$_{3}$ film with the lateral dimension of $10\times 5$~mm$^{2}$. (f) The liberated film was transferred to a target substrate (Si). (g) The PMMA layer was dissolved in acetone. (h) The Se capping layer was removed by annealing in the MBE chamber.}
  \label{Fig1}
\end{figure}

Figure~\ref{Fig1} illustrates the step-by-step protocol of liberating and transferring Bi$_{2}$Se$_{3}$ films with a lateral dimension of $10\times 5$~mm$^{2}$. Bi$_{2}$Se$_{3}$ is first grown on an oxide substrate, SrTiO$_{3}$~(111), using MBE (Figure~\ref{Fig1}a). The growth recipe is described in detail in the Methods section. Selenium capping after growth (Figure~\ref{Fig1}b) is crucial for protecting the film’s surface morphology against contamination from the polymer and the etchant used in later steps. Subsequently, the film is taken out of ultrahigh vacuum and a layer of polymethyl methacrylate (PMMA) is spin-coated on top of the selenium-capped film (Figure~\ref{Fig1}c). The PMMA layer is cured after baking at $80$~$^{\circ}$C for 10 minutes. The sample is then floated on buffered oxide etchant (BOE, HF:NH$_4$F=7:1) where the film peels itself off due to etching reactions at the film-substrate interface (Figure~\ref{Fig1}d,e). The floated film can be fished out using any desired substrate or device (Figure~\ref{Fig1}f). To facilitate ARPES measurements, the PMMA and Se layers are removed using acetone and 200 $^{\circ}$C vacuum annealing, respectively (Figure~\ref{Fig1}g,h).

\begin{figure}
\includegraphics[width=1\textwidth]{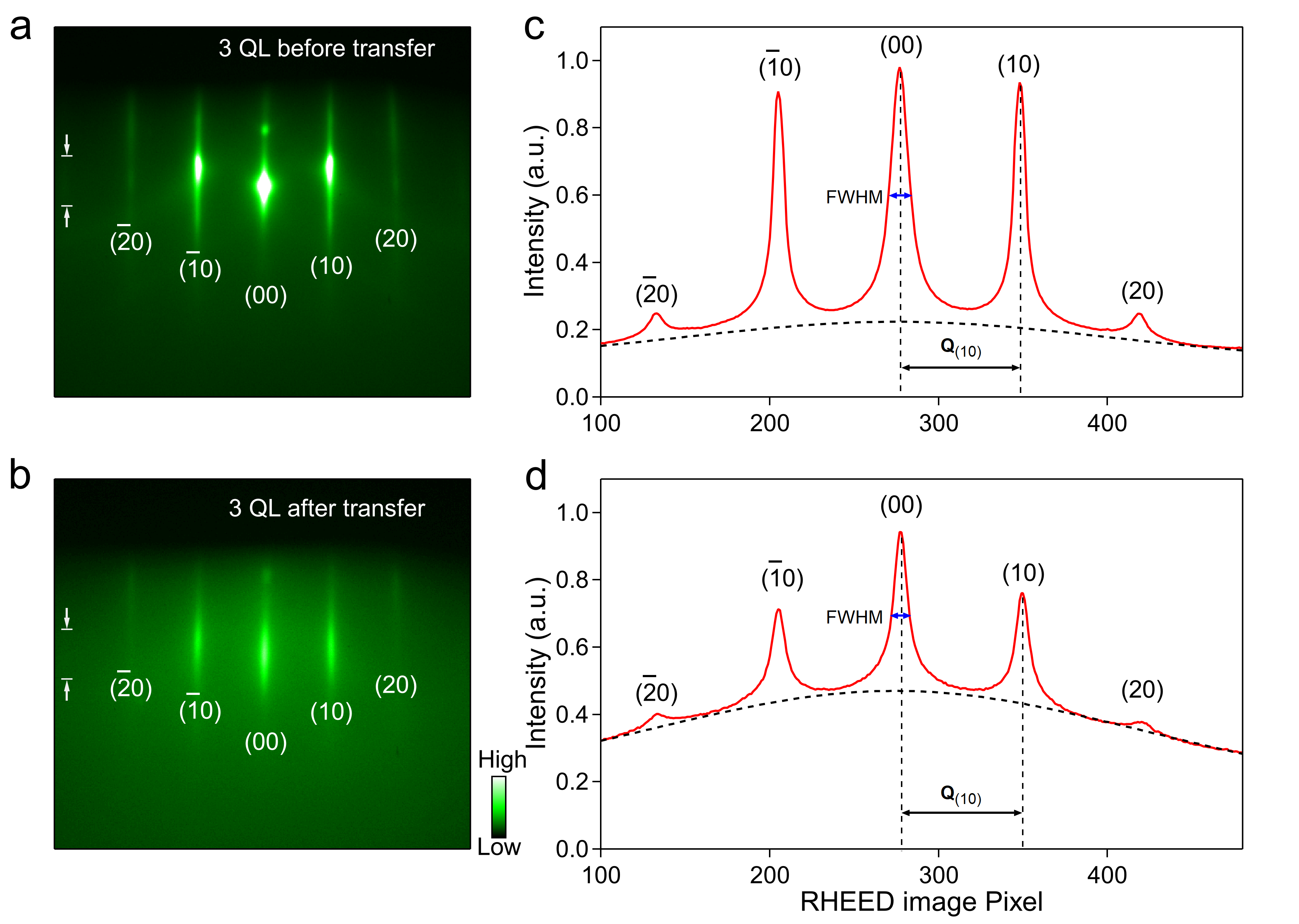}
  
  \caption{Reflection High Energy Electron Diffraction (RHEED) patterns of Bi$_2$Se$_3$ films taken with the electron beam along the [11$\overline{2}$0] direction. (a) RHEED image of an as-grown $3$~QL Bi$_2$Se$_3$ film on SrTiO$_3$ substrate. (b) RHEED image of a transfered $3$~QL Bi$_2$Se$_3$ film on Si. (c, d) The line profiles (red lines) of the RHEED images in (a) and (b), respectively. The integration ranges of the line profiles are marked by white arrows in (a) and (b). The black dashed thick lines are the backgrounds extracted by fitting the profiles using Gaussian functions. The signal-to-background ratios at the (00) peaks in (c) and (d) are around 3 and 1, respectively. The full-width-half-maximum (FWHM) of the (00) peaks and the distance between the (00) and (10) peaks are marked by blue and black arrows, respectively.}
  \label{Fig2}
\end{figure}

We perform reflection high energy electron diffraction (RHEED) characterization of the films to compare the surface morphology before and after the transfer. During the growth, the RHEED intensity at the specular peak undergoes strong oscillations (Supporting Information (SI) Note~1), allowing us to precisely determine the thickness. The RHEED image taken from an as-grown $3$~QL Bi$_2$Se$_3$ film shows streaks with a typical full-width-half-maximum (FWHM) of 13 pixels, which corresponds to 18\% of the distance between the zeroth and first-order streaks in the reciprocal space (Figure~\ref{Fig2}c). The signal-to-background ratio is near 3, as measured by the intensity ratio of the [00] streak and its baseline at the peak position. These observations indicate the two dimensionality and high crystallinity of the film. After the millimeter-scale transfer process, the RHEED pattern of the post-transfer film (Figure~\ref{Fig2}b) is highly consistent with the as-grown film (Figure~\ref{Fig2}a). The average FWHM of 12 pixels is comparable to the result from the as-grown film, while the signal-to-background ratio is reduced to 1 (Figure~\ref{Fig2}d). This is possibly due to the residue resulting from the polymer and other chemicals involved in the transfer process. Despite Selenium capping, a small amount of surface oxidation may also contribute to the background in the RHEED image. Nevertheless, the sharp streaks remain visible, suggesting that the surface morphology and crystallinity of the film are largely preserved after transfer. This is corroborated by atomic force microscopy (AFM) and Raman spectroscopy measurements on the transferred Bi$_{2}$Se$_{3}$ films (SI Note~2).

\begin{figure}
\includegraphics[width=1\textwidth]{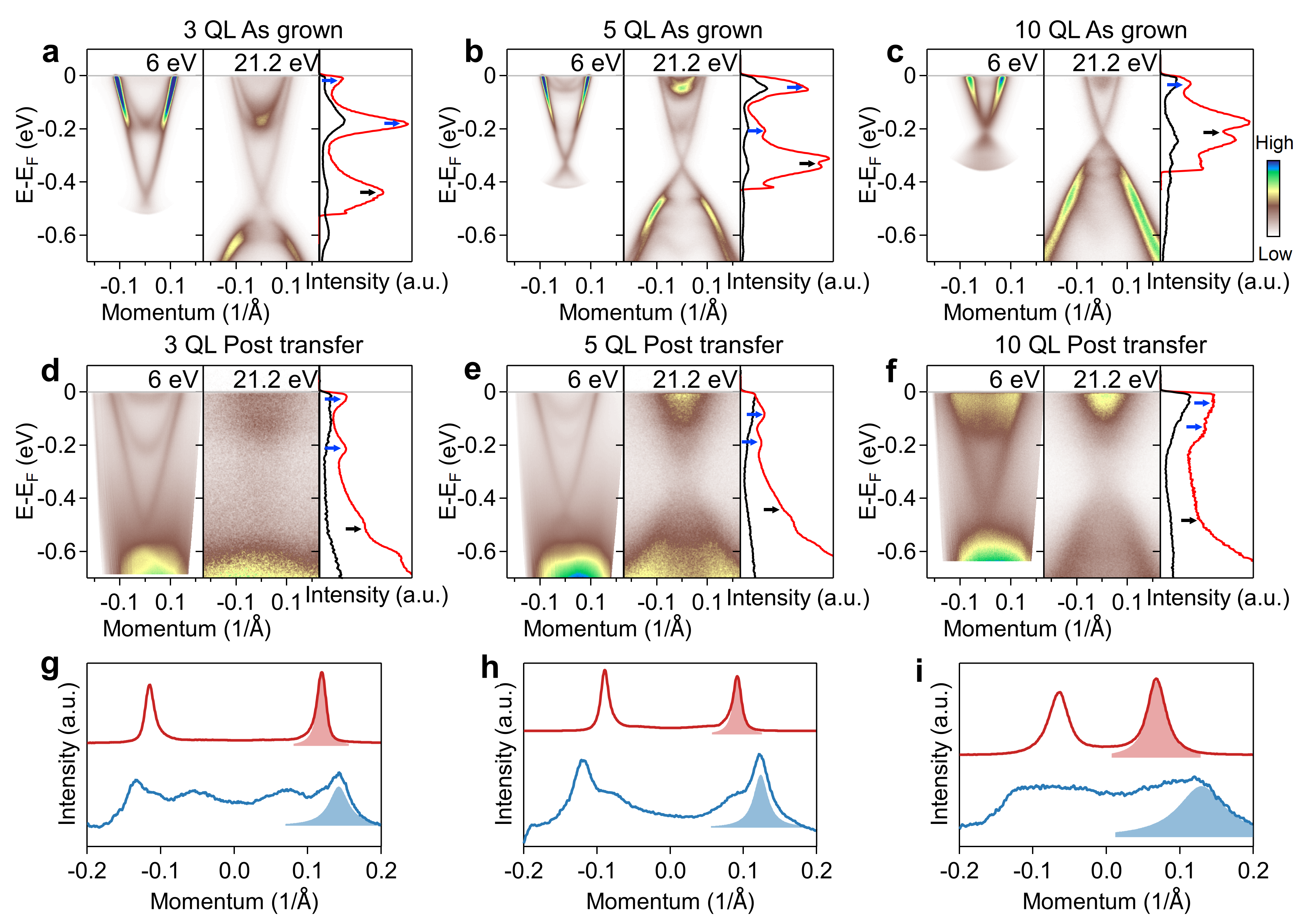}
  
  \caption{ARPES spectra of the as-grown and transferred Bi$_2$Se$_3$ films along the $\Gamma$-$K$ direction. (a-c) Band structures near $\Gamma$ of as-grown (a) $3$~QL, (b) $5$~QL, and (c) $10$~QL Bi$_2$Se$_3$ films. (d-f) Band structures near $\Gamma$ of exfoliated (d) $3$~QL, (e) $5$~QL, and (f) $10$~QL Bi$_2$Se$_3$ films. ARPES spectra taken with $6$~eV and $21.2$~eV photons are compared in each of the panels (a-f). Red and black lines in (a-f) are the energy distribution curves (EDCs) taken at $\Gamma$ using photon energies of $6$~eV and $21.2$~eV, respectively. Black and blue arrows mark the positions of the Dirac point and the quantum well states, respectively. (g-i) Momentum distribution curves (MDCs) taken at the Fermi level of the as-grown (red) and transferred (blue) Bi$_2$Se$_3$ films, using $6$~eV photons. Red and blue shaded areas illustrate the Lorentzian functions used to fit the peaks corresponding to the topological surface states.}
  
  \label{Fig3}
\end{figure}

We utilized ARPES to directly measure the electronic structures of the Bi$_2$Se$_3$ films before and after the transfer. Figures~\ref{Fig3}a-c show the band structures of the as-grown Bi$_2$Se$_3$ films with different thicknesses at $8.5$~K. Results taken with $6$~eV and $21.2$~eV photons are compared. Notably, low-energy cutoff's (LECs) are observed at $-0.52$~eV, $-0.42$~eV, and $-0.36$~eV for $3$~QL, $5$~QL, and $10$~QL films, respectively, in the $6$~eV ARPES spectra. The kinetic energy corresponding to the LEC is precisely the work function difference between the sample and the photoemission detector (SI Note~3)~\cite{Sobota2021RMP}. Using $21.2$~eV photoemission, the kinetic energies for the electrons near the Fermi level ($E_{\rm F}$) are in the range of $16$-$17$~eV, not affected by the LEC. The TSSs with quasi-linear band dispersions are clearly observed. Compared to a previous study on ultrathin Bi$_{2}$Se$_{3}$ films~\cite{QKXue2010YZhang}, our ARPES data using the same photon energy of $21.2$~eV feature sharper TSS band dispersions (Figure~\ref{Fig3}a-c, Figure~S4). Notably, when the thickness is $\leq 3$~QL the TSSs from the top and bottom surfaces are expected to hybridize~\cite{QKXue2010YZhang}. To avoid the complication due to the LEC, we use $21.2$~eV photons and resolve a subtle hybridization gap of $\sim 60$~meV near $E - E_{\rm F} \sim -0.52$~eV in the $3$~QL film (Figure~S3d), and a substantial gap of $\sim 200$~meV in the $2$~QL film (Figure~S4b). For thicknesses $\geq 5$~QL, $40$-$60$~meV gaps are seemingly observed at the Dirac point (Figure~\ref{Fig3}b, \ref{Fig3}c, Figure~S5), whereas this Dirac gap is expected to be $< 4$~meV by first principles calculations~\cite{liu_oscillatory_2010}. With a careful analysis of the thickness- and photon-energy-dependence of the Dirac gaps (SI Note 4), we conclude that the apparent Dirac gaps in $5$~QL and $10$~QL films before the transfer are likely due to the linewidth broadening effect, where the energy and momentum smearing leads to an apparent double-peak structure in the energy distribution curve at zero momentum (Figure~\ref{Fig3}b,c)~\cite{YLChen2019YJChen}, not reflecting the intrinsic physics in the materials. The Dirac point is systematically shifted to deeper binding energies as the film becomes thinner (Figure~\ref{Fig3}a-c), which is also consistent with previous observations\cite{QKXue2010YZhang}. This is likely due to the band bending arising from the charge transfer between the substrate and the Bi$_2$Se$_3$ film.  

In addition, QWSs induced by the quantum confinement are particularly clear in the $3$~QL and $5$~QL films (Figure~\ref{Fig3}a,b)\cite{QKXue2010YZhang}. We performed a systematic thickness-dependent study of the QWSs (SI Note~5). Theoretically, the energy difference between the $n^{th}$- and $0^{th}$-order QWSs ($\Delta E_{n-0}$) for an infinite-depth quantum well scales with $n^{2}/d^{2}$, where $d$ is the film thickness~\cite{liu_oscillatory_2010}. In our experiment, we extract the exponent of the $\Delta E_{1-0}$-versus-$d$ relationship to be $-1.75$ (Figure~S7), which is closer to the theoretical exponent of $-2$ as compared to $-0.78$ from the previous study~\cite{QKXue2010YZhang}. This result demonstrates that our as-grown Bi$_{2}$Se$_{3}$ films are of high quality and agree better with the theoretical model~\cite{liu_oscillatory_2010}.

Figures~\ref{Fig3}d-f display the ARPES spectra of the Bi$_{2}$Se$_{3}$ films after the transfer using both 6 eV and 21.2 eV photons. Our $6$~eV ARPES resolves clear electronic band structures from these millimeter-scale-transferred MBE-grown films. In these spectra, the TSS is preserved after the transfer for all three representative thicknesses. The LEC is out of the energy range of interest due to the reduced sample work function after the transfer (Figure~S3). The TSS exhibits an overall higher electron doping after the transfer. This can be understood as the loss of Selenium due to either its intrinsic volatility \cite{Analytis2010NP} or the chemical reaction of Bi$_2$Se$_3$ with H$_2$ \cite{PHofmann2010MBianchi} or H$_2$O \cite{RChristian2011HMBenia, XJZhou2012CYChen}. The Lorentzian linewidths of the TSSs extracted from the momentum distribution curves (MDCs) taken at $E\rm_F$ are listed in Table~\ref{Table1}. The linewidths from the post-transfer films are all less than three times the original linewidths from the as-grown films. Given the direct exposure of the Bi$_2$Se$_3$ films to BOE, deionized water, and air, the amount of broadening is relatively low, particularly for the $5$~QL film. This is in general consistent with the previous Hall measurements on post-transfer Bi$_2$Se$_3$ films \cite{NBansal2014}, which reported that the carrier mobilities from the post-transfer films can be comparable to or even higher than those from the pre-transfer films. We emphasize that the quantitative connection between the MDC linewidth and the carrier mobility requires more intricate theoretical treatment due to the inequivalence of the spectroscopic scattering rate and the transport scattering rate \cite{RGMani1988, Valla2000PRL}. The QWSs are also preserved in the $6$~eV ARPES spectra, yet the inter-band spacings $\Delta E_{n-0}$ are substantially modified, which is likely due to the change of the quantum well potential originated from the chemically induced surface and interface contamination (SI Note~6).

\begin{table}[ht]

\caption{Comparison of Lorentzian linewidths of the topological surface states extracted from the momentum distribution curves taken at $E_{\rm F}$.}
\begin{tabular}{c c c c}
\hline\hline
Thickness & FWHM as grown (\AA$^{-1}$) & FWHM post-transfer (\AA$^{-1}$) & Ratio of broadening \\
\hline
$3$~QL & 0.014 & 0.032 & 2.3 \\
$5$~QL & 0.013 & 0.023 & 1.8 \\
$10$~QL & 0.029 & 0.078 & 2.7 \\
\hline
\end{tabular}
\label{Table1}
\end{table}

\begin{figure}
\includegraphics[width=1\textwidth]{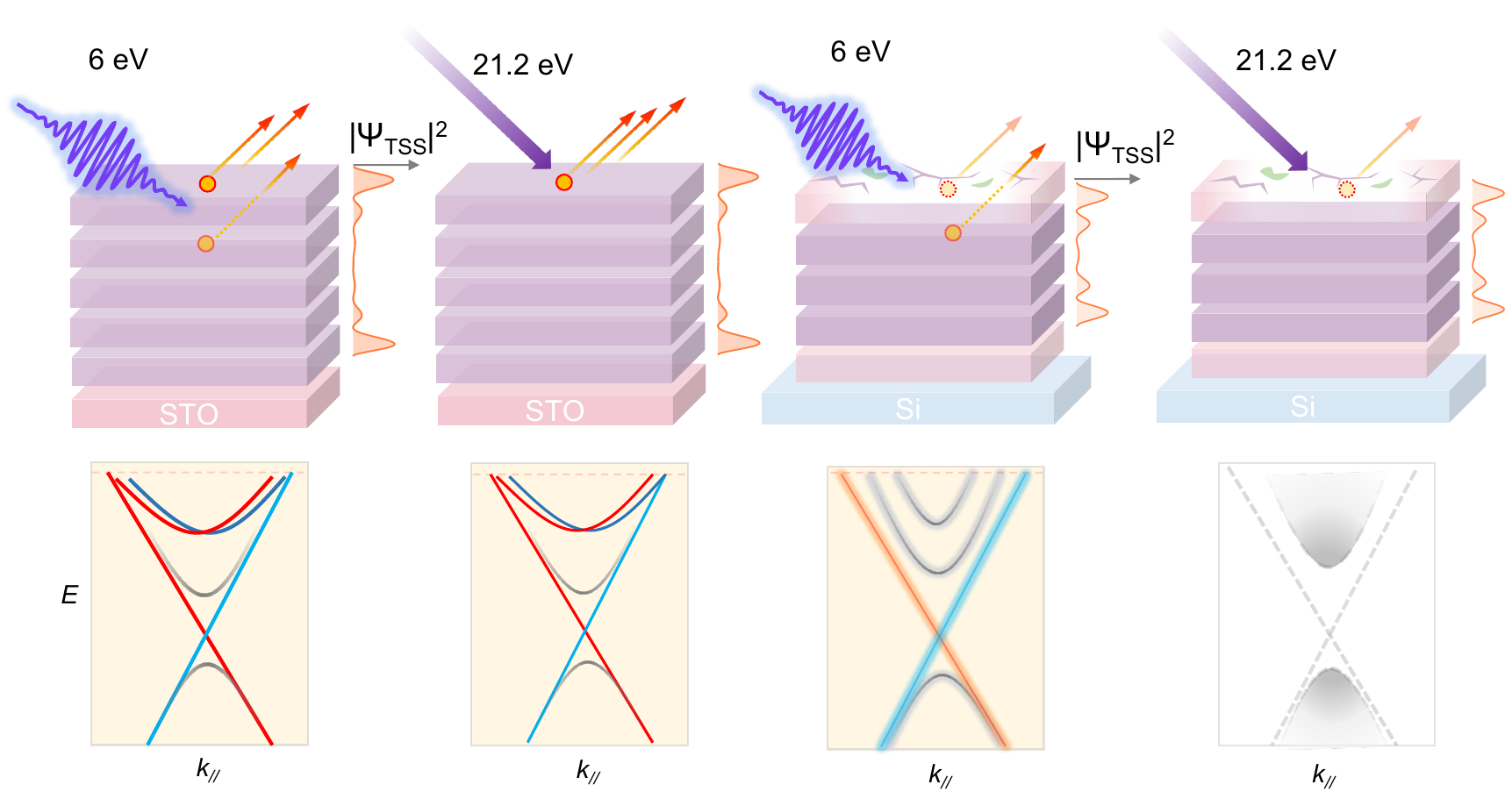}
  
  \caption{Wavefunction relocation due to transfer-induced contamination. Before the transfer, the topological surface state resides mostly in the outermost layers, and can be probed by both the surface-sensitive $21.2$~eV photoemission and the bulk-sensitive $6$~eV photoemission. After the transfer, the surface layer is disordered by interacting with O$_{2}$, H$_{2}$O, polymer, and Selenium, and the topological surface state is relocated partially into the inner layers. Only the $6$~eV photoemission can detect the electronic structure.}
  \label{Fig4}
\end{figure}

In contrast, the ARPES spectra on the post-transfer films taken with $21.2$~eV photons exhibit poorly defined bands. On the $5$~QL and $10$~QL films (Figure~\ref{Fig3}e,f), some features corresponding to the QWSs may still be recognizable, but with a substantial momentum smearing. The TSSs are clearly absent. Importantly, based on the universal mean free path of photoelectrons~\cite{seah1979}, $6$~eV photoemission has an effective probing depth of $3$-$4$~nm, while $21.2$~eV photoemission only detects the top $0.5$~nm. The different surface sensitivities help rationalize the photon energy dependence: the more surface-sensitive $21.2$~eV photoemission probes electrons primarily from the top QL which is significantly contaminated by O$_{2}$, polymers, Selenium, and other mechanisms; the more bulk-sensitive $6$~eV photoemission additionally reveals the electronic structures from the inner layers. By traversing the diffuse valence bands in the $5$~QL and $10$~QL films resolved by $21.2$~eV photons (Figure~\ref{Fig3}e,f), one can identify that the electronic structure for the surface layer is less electron-doped than that for the inner layers. This is consistent with the mechanism of electron-doping by the chemical reaction of Bi$_{2}$Se$_{3}$ and H$_{2}$O~\cite{RChristian2011HMBenia}, as the surface layer will be least impacted by the solution during the peeling-off process (Figure~\ref{Fig1}d, Figure~S8). Interestingly, the presence of the sharp TSSs in the $6$~eV data indicates at least a partial wavefunction relocation of the TSS from the top QL to the second QL. This microscopic understanding is consistent with a time-resolved ARPES study on MnBi$_{4}$Te$_{7}$ which unveiled a partial TSS wavefunction relocation due to the presence of surface defects~\cite{WJLeeNP2023}, but the effect is more drastic in the present case as a result of the heavy chemical processing. Moreover, we expect the TSS from the bottom QL to also undergo wavefunction relocation into the QL above it. The direct consequence of such wavefunction movements is an effective reduction of the number of QLs that comprise the TI (Figure~\ref{Fig4}), which is manifested in the opening of hybridization gaps in ultrathin films. Indeed, for the $3$~QL film we observe a $> 100$~meV gap at the Dirac point after the transfer (Figure~\ref{Fig3}d), which is significantly enhanced from the $\sim 60$~meV hybridization gap below the Dirac point before the transfer (Figure~\ref{Fig3}a, Figure~S3). We are unable to reliably extract the values of transfer-induced energy gaps from the $5$~QL or $10$~QL films (Figure~\ref{Fig3}e,f). Notably, no magnetic materials were used in the millimeter-scale transfer protocol, and therefore the magnetic impurities are unlikely to exist and cause the emerged Dirac gap. The transfer-induced gap can also be due to the hybridization between the TSS and a defect resonance state~\cite{XuYishuai2017}, but such a flat resonant state is not observed in our data for the $3$~QL film (Figure~\ref{Fig3}d). Hence, we attribute the transfer-induced Dirac gap on the $3$~QL film to the enhanced hybridization between the TSSs from the top and bottom layers due to the wavefunction relocation.

The transferability of our Bi$_{2}$Se$_{3}$ films allows us to create suspended $10$~QL membranes on top of a $5$~$\mu$m silicon nitride aperture. Although electron diffraction does not reveal measurable strains on selected areas of $200$~nm or larger (Figure~S9), atomic-resolution scanning transmission electron microscopy (STEM) resolves as much as $\pm 20$\% local strains concentrated near topological defects, which are formed before the transfer process and consistent with previous studies on substrate-supported Bi$_{2}$Se$_{3}$ films~\cite{LLi2014YLiu} (SI Note~7). An optical microscopy image is taken on a photonic hybrid system made by transferring a $6$~QL Bi$_{2}$Se$_{3}$ film onto an array of photonic crystal cavities (Figure~S10). With back lighting, this image highlights tearing lines which are separated by $5$-$20$~$\mu$m. These results not only demonstrate the realistic constraints on the size of the devices incorporating the suspended Bi$_{2}$Se$_{3}$ membranes, but also provide an explanation for our current inability to obtain ARPES results on these suspended membranes: additional oxidation and other channels of contamination can occur through the tears.


Our observation of the preservation of the TSS, together with the microscopic understanding of the TSS wavefunction relocation (Figure~\ref{Fig4}), provides an important physics foundation for the transferred ultrathin TI films. Even though wet-transfer methods~\cite{kim_remote_2017,NBansal2014} have greatly enhanced our capabilities to build artificial heterostructures with MBE-grown films, the chemical processing leads to concerns about the preservation of the electronic structures, in particular the surface- and interface-related electronic states. Note that for ultrathin TI films, a significant disorder in the material can indeed lead to a topology change since the topology is sensitive to where the wavefunction is relocated~\cite{liu_oscillatory_2010}. Therefore, it is far from trivial whether the TSS should be preserved in such a process. Our measurement, which reveals the preservation of the TSS for transferred Bi$_{2}$Se$_{3}$ films, paves the physics foundation for any future endeavors using the large-area transfer method for building heterostructures. Notably, this method can generally apply to other chalcogenide films grown on oxide substrates. In SI Note~8, we demonstrate success in millimeter-scale liberation and ARPES characterization on Bi$_2$Te$_3$ thin films. 

The microscopic understanding of the wavefunction relocation is also key to designing, rationalizing, and improving the functionalities of future TI-based devices. We note that the study pioneering the wet-transfer method for Bi$_{2}$Se$_{3}$ reported a trivial insulator behavior when tuning the chemical potential to the charge neutral point of a transferred $4$~QL film~\cite{NBansal2014}, which disagrees with the theoretical prediction of a quantum spin Hall insulator phase~\cite{liu_oscillatory_2010,QKXue2010YZhang}. Applying our new understanding of the TSS wavefunction relocation, the effective number of layers comprising a TI after the transfer can be $3$ or even smaller, which, according to the theoretical prediction~\cite{liu_oscillatory_2010}, can potentially make the system a trivial insulator. We emphasize that this understanding of wavefunction relocation will be particularly important in future endeavors of building stacked or twisted TI heterostructures~\cite{JCano2022DGuerci, JCano2022Dunbrack}, since the formation of Moir\'e bands sensitively depends on the wavefunction overlap at the constructed interface. In addition to the transferred films, wavefunction relocation also needs to be considered in devices fabricated using processing methods that could potentially degrade the interface, such as spin-orbit torque devices~\cite{JHan_2021} and TI-based Josephson junctions~\cite{Oliver2022}.

\section{Methods}
\subsection{Sample Growth.}
High-quality Bi$_{2}$Se$_{3}$ thin films were grown on 0.05wt\% Nb-doped  SrTiO$_{3}$ (111) substrates by molecular beam epitaxy (MBE). The base pressure of the MBE chamber was 2 $\times$ 10$^{-10}$~mbar. The SrTiO$_{3}$ (111) substrates were annealed at 1000$^{\circ}$C in the MBE chamber to obtain an atomically flat surface before growth. Selenium (99.999\% purity) and Bismuth (99.999\% purity) were co-evaporated onto the SrTiO$_{3}$ substrate from two effusion cells. The Bi$_{2}$Se$_{3}$ films were grown by using a two-step process \cite{SOh2011NBansal, JSHarris2013SHarrison}. One QL Bi$_{2}$Se$_{3}$ was first grown at a low substrate temperature of 225$^{\circ}$C. The rest layers were grown at a higher temperature of 265$^{\circ}$C. The amorphous Se capping layer was \textit{in situ} deposited on the film at room temperature. 

\subsection{STEM measurements.}
The atomic structural characterization was performed on an aberration-corrected scanning transmission electron microscope (STEM) by high-angle annular dark-field (HAADF) imaging using the aberration-corrected JEOL ARM200CF at the University of Illinois at Chicago. A cold field emission source operated at $200$~kV was equipped. The HAADF detector angle was 90-270 mrad to give Z contrast images with a less than $0.8$~Å spatial resolution. The $10$~QL Bi$_{2}$Se$_{3}$ film was transferred on top of a silicon nitride TEM grid with $5$~$\mu$m holes.

\subsection{AFM and Raman measurements.}
The AFM measurements were performed on a Bruker multimode 8 AFM at the University of Chicago. The Raman measurements were performed on a HORIBA LabRAM Microscope with a laser source of $532$~nm at the University of Chicago. A $6$~QL Bi$_{2}$Se$_{3}$ film was transferred on top of a silicon nitride TEM grid with $5$~$\mu$m holes. The Raman spectra were collected from the suspended Bi$_{2}$Se$_{3}$ film on the $5$~$\mu$m aperture at $300$~K.

\subsection{ARPES measurements}
High-resolution ARPES measurements were carried out on the University of Chicago's Multi-Resolution Photoemission Spectroscopy platform which is connected to the MBE chamber. $205$~nm photons were generated by frequency quadrupling of a Ti:Sapphire oscillator with an $80$~MHz repetition rate. The overall energy resolution was characterized to be better than $4$~meV\cite{SLYang2021CHYan}. The laser spot size characterized by the full width at half maximum (FWHM) was around $10$~$\mu$m.
\section{Author Contributions}
C.I.J.I. and Q.G. contributed equally to this work. C.I.J.I., Q.G., K.D.N., C.Y., W.L. and S.Y. grew the Bi$_{2}$Se$_{3}$ thin films. C.I.J.I., Q.G., K.D.N., C.Y., W.L., H.R., and S.Y. prepared the exfoliated Bi$_{2}$Se$_{3}$ thin films. C.I.J.I., Q.G., K.D.N., C.Y., W.L. and S.Y. conducted the ARPES measurements. G.Y., T.S.M., M.Z., Y.S.M., and C.L. conducted the TEM measurements. K.D.N., C.I.J.I., E.H., and C.Y. conducted the AFM measurements. E.H. and C.L. conducted the Raman measurements. Q.G., K.D.N., C.I.J.I., and S.Y. wrote the manuscript with input from all authors. S.Y. conceived this project.
\section{Acknowledgements}
The MBE and ARPES experiments were supported by DOE Basic Energy Sciences under Grant No. DE-SC0023317. The substrate preparation was partially supported by National Science Foundation under Award No. DMR-2011854. T.S.M. acknowledges support by the NSF Graduate Research Fellowship. This work used the JEOL ARM200CF in the Electron Microscopy Core, Research Resources Center at the University of Illinois at Chicago. 

\section{Supporting Information}
The Supporting Information is available free of charge via the internet at \href{http://pubs.acs.org}{http://pubs.acs.org}.

Layer-by-layer growth of Bi$_2$Se$_3$ films, AFM and Raman measurements of transferred Bi$_2$Se$_3$ films, low energy cutoff in the ARPES spectrum measured by $6$~eV, origins of the Dirac gaps in as-grown Bi$_2$Se$_3$ films, evolution of the quantum well states of Bi$_2$Se$_3$ films as a function of the film thickness, comparison of the quantum well states in Bi$_2$Se$_3$ films before and after the transfer, STEM characterization of suspended Bi$_2$Se$_3$ films, preservation of topological surface states in millimeter-scale transferred Bi$_2$Te$_3$ films.






\bibliography{BS}

\newpage

\newpage

  

\end{document}